\documentclass[structabstract,natbib]{aa}  
\usepackage{natbib}
\usepackage{graphicx}
\usepackage{txfonts}
\usepackage{array}
\usepackage{color}
\usepackage{rotating}
\usepackage{lscape}
\usepackage{multirow}
\usepackage{xcolor, soul}
\usepackage[colorlinks=true,linkcolor=blue,citecolor=blue,urlcolor=blue]{hyperref}

\bibpunct{(}{)}{;}{a}{}{,} 

\usepackage{amstext}

\begin{document}

  \title{An investigation of the star-forming main sequence considering the nebular continuum emission at low-$\mathrm{\it z}$}
  
  \author{Henrique Miranda\inst{1,2}, Ciro Pappalardo\inst{1,2}, Polychronis Papaderos\inst{1,2,3}, José Afonso\inst{1,2}, Israel Matute\inst{1,2}, Catarina Lobo\inst{3,4}, Ana Paulino-Afonso\inst{3}, Rodrigo Carvajal\inst{1,2}, Silvio Lorenzoni\inst{1}, Duarte Santos\inst{1,2}
  }
  \institute{Instituto de Astrof\'{i}sica e Ci\^{e}ncias do Espa\c{c}o, Universidade de Lisboa - OAL, Tapada da Ajuda, PT1349-018 Lisboa, Portugal
  \and
  Departamento de F\'{i}sica, Faculdade de Ci\^{e}ncias da Universidade de Lisboa, Edif\'{i}cio C8, Campo Grande, PT1749-016 Lisboa, Portugal
  \and
  Instituto de Astrof\'{i}sica e Ci\^{e}ncias do Espa\c{c}o, Universidade do Porto - CAUP, Rua das Estrelas, PT4150-762 Porto, Portugal
  \and
  Departamento de F\'{\i}sica e Astronomia, Faculdade de Ci\^encias, Universidade do Porto, Rua do Campo Alegre 687, PT4169-007 Porto, Portugal
  }


 
  \abstract
  {Galaxy evolution has been studied by interpreting the spectral energy distribution of galaxies using spectral synthesis codes. This method has been crucial in discovering different pillars of modern galaxy evolution theories. However, this analysis was mostly carried out using spectral synthesis codes that are purely stellar, that is, they assume that the nebular contribution to the total continuum is negligible. The code FADO is the first publicly available population spectral synthesis tool that treats the contribution from ionised gas to the observed emission self-consistently. This is expected to have a particularly strong effect in star-forming (SF) galaxies.}
  {We study the impact of the nebular contribution on the determination of the star formation rate (SFR), stellar mass, and consequent effect on the star-forming main sequence (SFMS) at low redshift.}
  {We applied FADO to the spectral database of the SDSS to derive the physical properties of galaxies. As a comparison, we used the data in the MPA-JHU catalogue, which contains the properties of SDSS galaxies derived without the nebular contribution. We selected a sample of SF galaxies with H$\alpha$ and H$\beta$ flux measurements, and we corrected the fluxes for the nebular extinction through the Balmer decrement. We then calculated the H$\alpha$ luminosity to estimate the SFR. Then, by combining the stellar mass and SFR estimates from FADO and MPA-JHU, the SFMS was obtained.}
  {The H$\alpha$ flux estimates are similar between FADO and MPA-JHU. Because the H$\alpha$ flux was used as tracer of the SFR, FADO and MPA-JHU agree in their SFR. The stellar mass estimates are slightly higher for FADO than for MPA-JHU on average. However, considering the uncertainties, the differences are negligible. With similar SFR and stellar mass estimates, the derived SFMS is also similar between FADO and MPA-JHU.}
  {Our results show that for SDSS normal SF galaxies, the additional modelling of the nebular contribution does not affect the retrieved fluxes and consequentially also does not influence SFR estimators based on the extinction-corrected H$\alpha$ luminosity. For the stellar masses, the results point to the same conclusion. These results are a consequence of the fact that the vast majority of normal SF galaxies in the SDSS have a low nebular contribution. However, the obtained agreement might only hold for local SF galaxies, but higher-redshift galaxies might show different physical properties when FADO is used. This would then be an effect of the expected increased nebular contribution.}

  \titlerunning{An investigation of the star-forming main sequence considering the nebular continuum emission at low-$\mathrm{\it z}$}
  \authorrunning{H. Miranda et al.}

  \maketitle
%

\section{Introduction}
\label{introduction}

Galaxies emit radiation that covers the whole range of the electromagnetic spectrum. The analysis of this radiation enables us to study their physical properties, formation, and evolution. In this context, a method that is widely used to extract the physical properties of galaxies is spectral modelling. The physical properties of extragalactic objects are retrieved and interpreted based on their spectral energy distribution (SED) through spectral synthesis methods. These methods were fundamental in the discovery of different pillars of modern galaxy evolution theories \citep[e.g.][]{kauff,brinch,trem,cid,gal,salim,man}.

One of the challenges of these methods is distinguishing all the different processes that contribute to the overall emission of the galaxy. This is related to the intrinsic nature of the observed light, which is a combination of emission from stars, gas, and dust. To fully model this emission, all these mechanisms must be included in a consistent framework. In the past, inverse population synthesis codes were built with the underlying assumption that the contribution to the observed emission in the optical waveband due to ionised gas was negligible. These codes are known as purely stellar codes. On the other hand, most evolutionary synthesis codes include the nebular emission, such as GALEV \citep{kot}, Pégase \citep{fioc}, and STARBURST99 \citep{leit}.

Most of the analysis of the spectroscopic data from one of the largest surveys ever made, The Sloan Digital Sky Survey \citep[SDSS;][]{SDSS}, has been carried out using inverse population synthesis codes. The SDSS provides one of the most complete databases for studies of galaxy evolution and has been fundamental in the investigation of this topic. Studies of the data obtained by the SDSS demonstrated that the distribution of galaxy properties is bimodal, which means that star-forming (SF) galaxies and passive galaxies have distinct properties \citep{gava}. The SDSS enabled the extensive measurement of fundamental galactic properties such as the star formation rate (SFR) and stellar mass \citep{brinch,kauff}. These two properties were found to be tightly correlated for SF galaxies, therefore this relation was called star-forming main sequence \citep[SFMS;][]{noe}. Since then, the SFMS has been extensively studied and was found to exist up to a redshift of 4, maintaining approximately the same slope, but increasing in normalisation factor \citep[see][for a compilation of results]{spea}. This is generally considered to be the fundamental relation of galaxy evolution theories today.

One particularly noteworthy spectroscopic analysis of the SDSS is provided by the Max Planck Institute for Astrophysics-Johns Hopkins University (MPA-JHU) catalogue, which analysed the spectroscopic data using a purely stellar approach. The currently available MPA-JHU catalogue\footnote{\url{https://wwwmpa.mpa-garching.mpg.de/SDSS/DR7}} provides a benchmark spectroscopic analysis of the Data Release 7 of SDSS \citep[SDSS-DR7;][]{SDSS_DR7}. The MPA-JHU catalogue has raw data with the object information (redshift, velocity dispersion, sky position, etc.) and spectroscopic measurements (line fluxes, equivalent widths (EWs), and continuum indices). The MPA-JHU catalogue has also derived metadata estimating SFRs, based on \citealt{brinch} (hereafter B04), and stellar masses, based on fits to the photometry following the philosophy of \cite{kauff} and \cite{salim}.

All these analyses of the SDSS spectroscopic data have been performed so far without considering the contribution of nebular gas. The modelling of the nebular gas has been neglected because it is complex to implement this in spectral synthesis codes and because the nebular component was expected to be low for SDSS galaxies. However, the contribution from the nebular emission, that is, the nebular gas surrounding SF regions, can rise up to 60\% of the total emission \citep{krue,izo,papad,leit,anders,scha09}. Hence, even if the assumption of negligible nebular emission might be reasonable for galaxies with low SFRs, it requires some caution when applied to galaxies with higher SFRs \citep{gus,cardo,pappa}. 

The inclusion of the nebular component in the analysis affects the estimation of the continuum. In the case of a purely stellar code, the continuum is fit considering it is only due to stellar contribution, whereas in FADO, the continuum is fit with a combination of the stellar and the nebular emissions. FADO considers the spectral shape of the continuum in the analysis, and this leads to a different parameter estimation. The essence of the problem is not primarily the total luminosity fraction of the nebular continuum, but its spectral shape: whereas the ionising stellar continuum is blue, the ionised gas (specifically, the nebular continuum) is reddish/flat (in the optical, excluding the Balmer and Paschen discontinuity). From the observational point of view, this was discussed in \cite{papad} (see their Fig. 8), who presented a spectral decomposition into the stellar and nebular continuum. Their figure and also Fig. 9 of \cite{izo11} shows that the relative contribution of the nebular continuum increases with wavelength. In other words, the nebular continuum makes the total SED redder than the pure SED of the stellar component. As pointed out in \cite{izo11}, this reddish continuum forces a purely stellar code to invoke a much too high fraction of red stellar populations (i.e. old and with a high mass-to-light ratio), consequently to overestimate the stellar mass. On the other hand, to compute SFRs based on emission line fluxes, the local continuum needs to be subtracted, which is also handled differently in different codes.

In face of these issues, and because it has been shown in particular that the contribution of the nebular component can be significant for some galaxies, it is fundamental to evaluate its effect on the relations between galactic properties that have been established throughout the years. This required a proper modelling of the stellar and nebular contribution to the observed emission of a galaxy.

The code FADO\footnote{\url{https://spectralsynthesis.org/fado}} \citep[fitting analysis using differential evolution optimization;][]{gom} is the first population spectral synthesis code capable of fitting the optical SED due to the stellar and nebular components self-consistently. The usage of FADO and its novel approach to SED fitting is expected to play a decisive role in shedding light into the impact of the nebular contribution when spectroscopic data are analysed and how this might have influenced our current understanding of the processes taking place in galaxies. The capabilities of FADO were tested in a series of benchmark publications \citep{cardo,pappa}.

\cite{cardo_prep} compared FADO with STARLIGHT \citep{cid} by applying these tools to a set of synthetic spectra of galaxies with different star formation histories (SFHs). The consideration of different SFHs allowed the comparison of the results between FADO and STARLIGHT as a function of the level of the nebular contribution. The stellar masses, light- and mass-weighted stellar ages, and metallicity estimates of the two tools were compared, and relevant differences were found. For galaxies with a significant nebular contribution, STARLIGHT can overestimate the stellar mass and mass-weighted mean stellar age up to $\sim$2 dex and $\sim$4 dex, respectively. On the other hand, STARLIGHT underestimates the mean metallicity and light-weighted
mean stellar age by up to $\sim$0.6 dex. Compared to these results, FADO obtains significantly better estimates because these properties can be estimated with a precision of $\sim$0.2 dex, even for evolutionary stages where the nebular contribution is highly prominent. \cite{pappa} followed a similar method to study the relation between the signal-to-noise ratio (S/N) and the accuracy of the estimated physical properties. Moreover, this work related the overestimation of the mass-weighted quantities by pure stellar codes to their difficulty in modelling the region of the spectrum around the Balmer jump.

The next step after these initial tests was to apply FADO to real spectra, and with this goal in mind, FADO was applied to the SDSS-DR7 \citep{cardo_prep}. \cite{cardo_prep} carried out a similar analysis to that of \cite{cardo}, but this time to real data. In Section \ref{discussion}, this work and the results obtained are discussed in more detail.

In this paper, we intend to follow the line of investigation of re-evaluating previously established relations between physical properties of galaxies in the light of the novel self-consistent approach of FADO, focusing on the SFMS. This is relevant because modelling the nebular continuum is expected to be particularly influential for SF galaxies, in which the nebular contribution is higher in general compared to other types of galaxies. Studying the SFMS with FADO in a well-known data set such as SDSS-DR7 is therefore expected to clarify the importance of modelling the nebular continuum for SF galaxies at low z. Moreover, it will serve as a reference test, providing insights for future studies of the SFMS with FADO at higher redshifts. This will help the current effort to understand the best tool for modelling data from future surveys at higher redshifts.

The aim of this paper is to quantify the impact of the nebular contribution on the estimation of galactic properties in SF galaxies. To do this, results from the application of FADO to the SDSS-DR7 are used to estimate the SFR and stellar mass. Next, we compare our results with a previous analysis of the SDSS-DR7, quantifying the differences due to the nebular contribution. In this way, the results from FADO are compared to the data in the MPA-JHU catalogue, which was obtained using a purely stellar approach. In summary, our objectives are a) to quantify and understand the influence of the nebular component in the determination of the SFR, stellar mass, and SFMS, abd b) to relate the obtained results with the contribution of the nebular emission to the total emission.

The paper is organised as follows: Section \ref{sample} provides an overall view of the SDSS and the selection criteria of the studied sample. Section \ref{method} details the spectral fitting tool FADO. In Sects. \ref{SFR_det} and \ref{Mstar_det} we explain how the SFR and stellar mass were derived, respectively. Section \ref{results} presents the results of the comparison of the SFR, stellar mass, and SFMS between FADO and MPA-JHU. In Section \ref{discussion} we discuss the results. Finally, Section \ref{conclusion} provides a summary of the main results of this work and their relevance. We use a cosmology with $H_{0}$ = 70 km s$^{-1}$ Mpc$^{-1}$, $\Omega_{M}$ = 0.3, and $\Omega_{\Lambda}$ = 0.7. We consider the \cite{chabr} initial mass function (IMF) to obtain the SFR and stellar mass estimates. To convert these into a \cite{krou} IMF, the estimates should be multiplied by 1.06 \citep{zah,spea,madau}.

\section{Sample}
\label{sample}

The SDSS is a survey of photometric and spectroscopic data across $\pi$ sr, conducted with the 2.5 m telescope at the Apache Point Observatory. The galaxies analysed in this study were drawn from the SDSS-DR7\footnote{\url{http://classic.sdss.org/dr7/}} \citep{SDSS_DR7}, which has spectroscopic data for a wavelength coverage of 3800 - 9200 \AA with a resolution of 1800 - 2200 and an S/N $>$ 4 per pixel at $g$ = 20.2 for 926 246 objects. The spectra were obtained with fibres that have a sky aperture of 3 arcsec. The SDSS-DR7 also has photometric data for these objects in five bands ($u$, $g$, $r$, $i$, $z$).

We considered the application of FADO to SDSS-DR7 carried out \cite{cardo_prep}\footnote{A repository with the data is being prepared and will be soon publicly available.}. We started by finding the galaxies in common in the FADO data and the galaxies available in the MPA-JHU catalogue. There are 916 118 in common galaxies in the two sets of data. To remove low flux values, the following conditions were defined for H$\alpha$ and H$\beta$ for both FADO and MPA-JHU: $-$20 $<$ $\log$(F[erg s$^{-1}$cm$^{-2}$]) and S/N $>$ 3 (680 432 galaxies were removed with these criteria). We set a lower mass limit of 10$^{5}$ M$_{\odot}$ for both FADO and MPA-JHU to ensure that no low-mass peculiar objects or massive globular clusters were included in the sample (19 517 galaxies were removed with this criterion). We removed these objects because they might have a non-linear relation between SFR and stellar mass. With the described constraints, the sample was reduced to 216 169 galaxies, approximately 23\% of the common sample. From this sample, only the SF galaxies were selected, as described next.  

FADO uses a different galaxy classification scheme with respect to MPA-JHU (for details, see \citealt{brinch}). For both FADO and MPA-JHU, the classification is based on the emission line properties of the galaxies following the \citet[][hereafter BPT]{BPT} diagram. FADO classifies the galaxies strictly using their position on the BPT diagram, leading to the following categories: Pure SF, SF, Composite, LINER, and Seyfert. On the other hand, MPA-JHU considers in addition to the position on the BPT diagram also the S/N of the BPT lines, resulting in the following categories: SF, low S/N SF, Composite, AGN non-LINER, and low S/N LINER. Moreover, for both FADO and MPA-JHU, galaxies that lack a flux measurement for one of the BPT lines are said to be Unclassifiable.

To select a common sample of SF galaxies, two subsamples were defined: SF galaxies, and non-SF galaxies. The subsample of SF galaxies for FADO includes Pure SF and SF galaxies, and for MPA-JHU, the sample includes SF and low S/N SF galaxies. The subsample of non-SF galaxies for FADO and MPA-JHU includes all the remaining categories except for the Unclassifiable galaxies. Finally, the galaxies equally classified as SF or non-SF in both FADO and MPA-JHU were selected, from which we obtained a sample of 182 933 SF galaxies and 25 220 non-SF galaxies. This information is summarised in \autoref{tab:gal_type_new_class}. The SF sample of 182 933 galaxies is the final sample used in this work.

\begin{figure}[h!]
        \centering
        \includegraphics[scale=0.5]{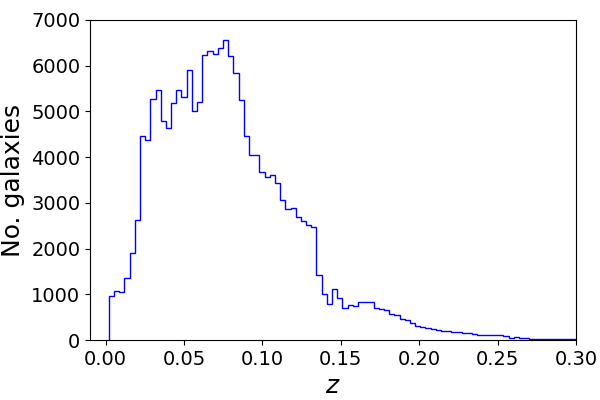}
        \caption{Redshift distribution of the galaxies in our sample. Total number of galaxies: 182 933.}
        \label{fig:z_distribution}
    \end{figure}

The mean redshift of our sample is $\langle z \rangle=0.07\pm0.04$. The redshift distribution of the galaxies is presented in \autoref{fig:z_distribution}.

    \begin{table*}[h!]
        \centering
        \caption{Galaxy classification according to the categories defined in this work. In the FADO and MPA-JHU columns, the total number of galaxies is 216 169, and for the last column, it is 208 153.}
        \label{tab:gal_type_new_class}
        \begin{tabular}{ccccccc}
        \hline
        & \multicolumn{2}{c}{FADO} & \multicolumn{2}{c}{MPA-JHU} & \multicolumn{2}{c}{FADO \& MPA-JHU} \\ \hline
        Galaxy type & Number & Percentage & Number & Percentage & Number & Percentage \\ \hline
        SF & 183 460 & 85 & 190 374 & 88 & 182 933 & 88 \\
        Non-SF & 31 184 & 14 & 25 787 & 12 & 25 220 & 12 \\
        Unclassifiable & 1 525 & 1 & 8 & $>$ 1 & 0 & 0 \\ \hline
        \end{tabular}
    \end{table*}

\section{Method}
\label{method}

FADO \citep{gom} is the first population spectral synthesis code that self-consistently fits the optical emission due to the stellar and nebular components. This allows a spectroscopic analysis that takes into account the contribution of the nebular gas emission to the observed SED, which is a novelty compared to previous spectral synthesis codes.

FADO employs a differential genetic optimisation approach known as differential evolution optimisation \citep[DEO;][]{sto96}. The benefit of this approach is a convergence to solutions in a stable way within a reasonable computational time. Another innovation of FADO is the consistent treatment of the main nebular characteristics, in particular, the luminosities and EWs of H$\alpha$ and H$\beta$ and the continuum near the Balmer and Paschen jumps when determining the physical and evolutionary properties of SF galaxies. The fitting algorithm is responsible for this consistent treatment of the nebular component, ensuring that during the minimisation process, the various stellar components are consistent with the expected line and continuum nebular emission characteristics. For instance, at the end of the convergence procedure, the production of ionising radiation by the stellar component is consistent with the nebular continuum observed and the Balmer line fluxes. This is possible because the estimation and inclusion of the nebular continuum in the best-fitting SED and the calculation of the electron density and temperature of the ionised gas is made on the fly.

The model SED for FADO, $F_{\lambda}$, results from a linear combination of $N_{*}$ spectral components dependent on the considered set of base spectra. The model SED has a stellar component to which the emission from stars contributes, plus a nebular component that is due to the nebular continuum and the hydrogen recombination lines. This model SED is given by \citep{gom}

        \begin{equation}
            \label{eq:FADO}
            \begin{aligned}
                F_{\lambda} &= \overbrace{\sum_{j=1}^{N_{*}} M_{j,\lambda_{0}} L_{j,\lambda_{0}} \times 10^{-0.4 A_{v} q_{\lambda}} \otimes S \left( v_{*}, \sigma_{*} \right) }^{Stellar\,Term} \\
                &+ \overbrace{ \Gamma_{\lambda} \left( n_{e}, T_{e} \right) \times 10^{-0.4 A_{v}^{neb} q_{\lambda}} \otimes N ( v_{\eta}, \sigma_{\eta}) }^{Nebular\,Term}
            \end{aligned}
        \end{equation}
        
\noindent On the stellar term, $M_{j,\lambda_{0}}$ is the stellar mass of the jth spectral component at the normalisation wavelength $\lambda_{0}$, $L_{j,\lambda}$ is the spectrum of the jth element, $A_{v}$ is the V band extinction, and $S \left( v_{*}, \sigma_{*} \right)$ represents the stellar kinematics. On the nebular term, $\Gamma_{\lambda} \left( n_{e}, T_{e} \right)$ is the nebular continuum spectrum where $n_{e}$ and $T_{e}$ are the electron density and temperature, respectively, $A_{v}^{neb}$ is the gas extinction, and $N(v_{\eta}, \sigma_{\eta})$ represents the nebular kinematics. In both terms, $q_{\lambda}$ is the ratio of $A_{\lambda}$ over the V-band extinction.

The self-consistency of FADO means that both terms of Eq. \ref{eq:FADO} are calculated simultaneously while ensuring their consistency. Pure stellar codes only provide information regarding the first term, requiring the use of photoionisation codes to model the nebular contribution in a second step of the analysis. This separate treatment of the stellar and nebular components leads to a loss of consistency between the two components. It should be noted that there have been other attempts at a consistent fitting of the emission from gas and stars using different approaches \citep[e.g.][]{gus01,acqua,amorin,cheval}. However, these approaches did not strictly follow the principles of population spectral synthesis. For example, \cite{gus01} and \cite{amorin} only considered old and young SSPs instead of a full library of SSPs. In summary, the main aspects of FADO are i) that both the stellar and nebular continuum are included in the fit, ii) the emission lines are automatically masked, iii) the extinction law is set as an input parameter, and iv) instrumental and physical broadening effects are included.

The parameter space is explored and the minimum $\chi^{2}$ between the observed and fitted SED is searched for while ensuring that the various stellar components are consistent with the expected line and continuum nebular emission characteristics. This consistency in the fitting of the stellar and nebular components ensures the previously mentioned self-consistency between the stellar and nebular optical emission.

The application of FADO to the SDSS-DR7 spectroscopic data was fully described in \cite{cardo_prep}, and the spectrum pipeline analysis used to derive the data in the MPA-JHU catalogue was presented in \cite{trem}. Here, we only summarise the procedures adopted in these two works.

\cite{cardo_prep} corrected the raw SDSS-DR7 spectra for the foreground extinction considering the dust maps from \cite{schleg} and were then converted into the observer restframe. Next, the spectra were fitted with FADO in the wavelength range between 3400 and 8900 \AA\,with a base library from \cite{bruz03} composed of 150 simple stellar populations (SSPs), containing 25 ages (between 1 Myr and 15 Gyr) and six metallicities (1/200, 1/50, 1/5, 2/5, 1, and 2.5 Z$_{\odot}$). A \cite{chabr} IMF was considered to match the base library, Padova 1994 evolutionary tracks \citep{alon,bress,fagoa,fagob,girar}, and the \cite{calz00} extinction law.

To derive the data in the MPA-JHU catalogue, the continua of the spectra were corrected for the foreground extinction following the \cite{schleg} dust maps and were then fitted in the wavelength range between 3200 and 9300 \AA\,following the models by \cite{bruz03} with ten ages (between 5 Myr and 10 Gyr) and three metallicities (1/5, 1 and 2.5 Z$_{\odot}$). A \cite{krou} IMF, Padova 1994 evolutionary tracks \citep{alon,bress,fagoa,fagob,girar}, and the \cite{char_fall} extinction law were used.

There are differences between the two approaches that affect the derived results, in particular the IMF and extinction law. A different IMF impacts the SFR and stellar mass estimates. However, this impact can be straightforwardly solved by applying a conversion factor between the estimates obtained with the two IMFs. In the case of the extinction law, \cite{dicki} and \cite{papo} suggested that the impact of using different extinction laws is negligible. Thus, the fact that different extinction laws are used in the spectral modelling should not play a significant role in the derived results. The differences in the age and metallicity are not expected to have a significant impact on the final estimations because a different choice of SSP ages will mostly affect the star formation history, while a different range of metallicity does not change the stellar mass estimates for galaxies at $z$ < 1, as discussed in \cite{spea}.

Moreover, the fluxes in the MPA-JHU catalogue were renormalised to match the photometric fiber magnitude in the \textit{r} band, which is a different spectrophotometric calibration than the one considered in SDSS-DR7. This renormalisation factor is provided in the MPA-JHU catalogue and was applied to the flux and mass estimates used in this work so that they are comparable between the two data sets.

\section{SFR determination}
\label{SFR_det}

We focused on SF galaxies. A common tracer of the SFR used in the literature is the H$\alpha$ flux \citep[e.g.][]{galla,kenn,shi,zah,sob}, and it was also the tracer we used. The relation between the H$\alpha$ flux and the SFR is discussed in Sect. \ref{SFR_Halpha_flux}.

\subsection{Emission line properties}
\label{lines}

Prior to computing the SFR, it should be insightful to characterise and compare the properties of the emission lines retrieved by FADO and MPA-JHU. To begin, it is important to have a sense of the relation between the observed flux of an emission line characteristic of SF regions for FADO and MPA-JHU. This information is a first indicator of the relation between the estimated SFR for FADO and MPA-JHU. 

\begin{figure}[h!]
        \centering
        \includegraphics[scale=0.35]{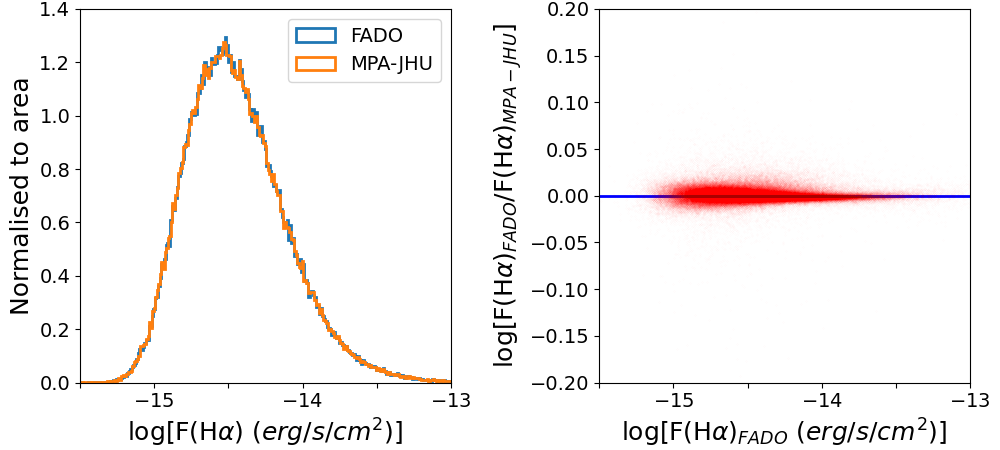}
        \caption{Comparison of the observed H$\alpha$ flux between FADO and MPA-JHU. Left panel: Histogram of $\log$[F(H$\alpha$)] for FADO (blue) and for MPA-JHU (orange). Right panel: Scatter plot of the logarithm of the ratio of the H$\alpha$ flux of FADO and MPA-JHU as a function of $\log$[F(H$\alpha$)] for FADO.}
        \label{fig:observed_fluxes}
    \end{figure}

 \autoref{fig:observed_fluxes} compares the observed flux of H$\alpha$  between FADO and MPA-JHU. The H$\alpha$ flux distributions are consistent, showing negligible differences. This indicates a similar SFR estimate.

\begin{figure}[h!]
        \centering
        \includegraphics[scale=0.35]{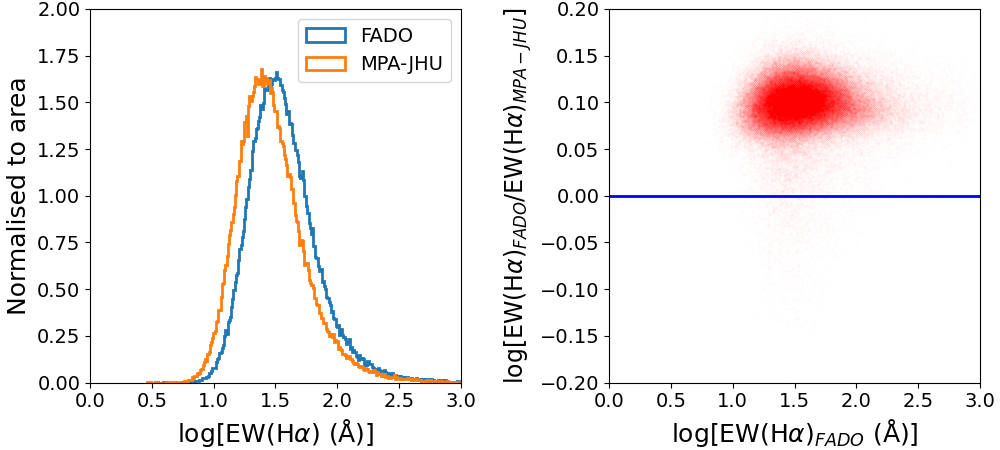}
        \caption{Comparison of the observed H$\alpha$ equivalent width for FADO and MPA-JHU. Left panel: Histogram of $\log$[EW(H$\alpha$)] for FADO (blue) and for MPA-JHU (orange). Right panel: Scatter plot of the logarithm of the ratio of the  H$\alpha$ EW of FADO and MPA-JHU as a function of $\log$[EW(H$\alpha$)] for FADO.}
        \label{fig:observed_EW}
    \end{figure}

In addition to the flux, another important characteristic of a spectral line is the EW. \autoref{fig:observed_EW} compares the EW of H$\alpha$  between FADO and MPA-JHU.

The peak of the distribution for FADO corresponds to higher EW values than the peak for MPA-JHU. The right panel of \autoref{fig:observed_EW} clearly shows that FADO obtains higher EWs than MPA-JHU. Almost all the galaxies are distributed above the blue reference line which marks equality between the estimates. This means that for practically all SF galaxies, FADO retrieves higher EWs than MPA-JHU. Their median difference shows that FADO estimates higher values than MPA-JHU by 0.1 dex ($\approx$26\%).

FADO measures higher EWs because it considers a different definition of the level of the continuum. For FADO, the continuum is the monochromatic continuum level at $\lambda_{0}$, which takes into account possible absorption features and stellar velocity dispersion. On the other hand, for MPA-JHU, the continuum is obtained by averaging the level of the continuum adjacent to the emission or absorption line. The latter approach is most commonly used in the literature. Because the stellar absorption is taken into account by FADO, the EW values are higher than in MPA-JHU.

A consequence of the alternative (but more physically motivated) definition of the continuum level in FADO is the observed measurement of larger H$\alpha$ EWs than in MPA-JHU. This difference in the definition of the continuum is only relevant for recombination lines, such as H$\alpha$, because of the absorption feature below the emission line. For forbidden lines, such as [OIII]5007, both continuum definitions lead to compatible EWs.

\begin{figure}[h!]
        \centering
        \includegraphics[scale=0.55]{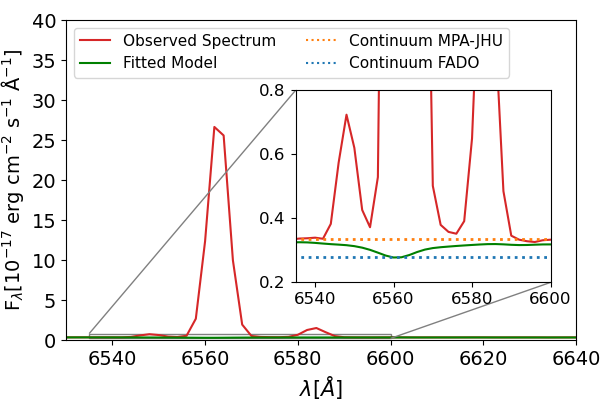}
        \caption{Comparison of the definition of the continuum for FADO and MPA-JHU for the H$\alpha$ line. The red line is the observed spectrum, and the green line is the fitted model. The dashed orange line represents the continuum as defined from averaging the level of the continuum adjacent to the emission line (MPA-JHU definition). The dashed blue line represents the continuum considering the bottom of the absorption feature below the emission line (FADO definition).}
        \label{fig:EW_plot_comparison}
    \end{figure}

As explained, the level of the continuum from which the emission line properties are measured is different between FADO and MPA-JHU. From the definition of EW, the level of the continuum is given by the ratio of the flux and the EW. Thus, because FADO obtains in median H$\alpha$ EWs 0.1 dex higher than MPA-JHU, then the continuum is in median 0.1 dex higher for MPA-JHU than for FADO. \autoref{fig:EW_plot_comparison}  shows the different definitions of the continuum for FADO and MPA-JHU for the H$\alpha$ emission line.

Before using the H$\alpha$ flux to derive the SFR, we corrected the flux for the intrinsic extinction caused by the gas and dust in the line of sight of the observer. The flux was corrected for the nebular extinction through the Balmer decrement (H$\alpha$/H$\beta$) and using the extinction law from \cite{char_fall} to be consistent with the procedure from MPA-JHU. The procedure is fully described in \autoref{appendix}.

\subsection{SFR from the H$\alpha$ flux}
\label{SFR_Halpha_flux}

Young and massive stars emit a large amount of ionising radiation that ionises the atoms in the surrounding gas, thus producing emission lines. In this way, strong emission lines are characteristic of SF regions, in particular, the lines from the Balmer series such as H$\alpha$ and H$\beta$. Because the most significant contribution to the ionising flux comes from stars with masses higher than 10 M$_{\odot}$ and lifetimes short than 20 Myr, the emission lines are a tracer of the current SFR \citep{kenn}. The H$\alpha$ line is a strong line and is located in the optical wavelength range. It is therefore typically used as a tracer of the current SFR. 

The H$\alpha$ luminosity can be calculated from the H$\alpha$ flux, and these two quantities are related by

    \begin{equation}
        L(H\alpha) = F(H\alpha)4\pi d_{L}^{2}
        \label{eq:Lum_to_flux}
    ,\end{equation}
    
\noindent where L(H$\alpha$) and F(H$\alpha$) are the H$\alpha$ luminosity and flux, respectively, and $d_{L}$ is the luminosity distance. The SFR and the H$\alpha$ luminosity are related by

    \begin{equation}
        \textrm{SFR} = \dfrac{L(H\alpha)}{\eta(H\alpha)}
        \label{eq:SFR}
    ,\end{equation}

\noindent where L(H$\alpha$) is the H$\alpha$ luminosity and $\eta$(H$\alpha$) is the conversion factor to the SFR. B04 used the \cite{kenn} conversion factor $\eta$(H$\alpha$)=10$^{\,41.28}$ erg s$^{-1}$M$_{\odot}^{-1}$yr. For consistency with B04, we assumed the same conversion factor, but calibrated to a \cite{chabr} IMF, $\eta$(H$\alpha$)=10$^{\,41.31}$ erg s$^{-1}$M$_{\odot}^{-1}$yr.

To derive a conversion factor between SFR and L(H$\alpha$), a set of assumptions were considered, such as a continuous star formation over a period of about 100 Myr and a metallicity equal to solar abundances. \cite{weilb} showed that deviations from these assumptions can lead to significant errors in the estimated SFR for particular types of galaxies, such as star-bursting dwarf galaxies. However, when our sample selection criteria are taken into account, galaxies with significant deviations from the mentioned assumptions, such as starburst galaxies, represent a very small fraction of the galaxies in our sample. Additionally, because we derived the SFR through the same procedure for FADO and MPA-JHU both, the concerns raised in \cite{weilb} do not affect our goal of comparing SFR estimates between FADO and MPA-JHU.

In summary, with \autoref{eq:Lum_to_flux}, the H$\alpha$ flux can be converted into H$\alpha$ luminosity. With \autoref{eq:SFR} ,  the SFR can then be estimated.

\subsection{Conversion from fibre SFR into total SFR}

The SDSS is a fibre-based survey. In this way, the SFR derived from the flux corresponds only to the fibre SFR (hereafter, referred to as SFR$_{fib}$), that is, the SFR of the galactic region sampled by the fibre. The fibre only samples $\text{about one-third}$ of the total galaxy light at the median redshift of the SDSS \citep{brinch}. Hence, when the total SFR (SFR$_{tot}$) is to be obtained, the SFR$_{fib}$ needs to be corrected for aperture effects.

B04 corrected the SFR estimates for aperture effects. However, \cite{salim} showed that the method used in B04 overestimated the SFR for galaxies with low SFRs. They also reported that it was possible to remove this bias. To improve the aperture correction, the light outside the fibre was calculated for each galaxy, and then stochastic models similar to those used by \cite{salim} were fitted to this photometry. These improved SFR$_{tot}$ estimates are available on the MPA-JHU website.

There have been more recent attempts to empirically correct fiber-based SFRs for aperture effects, taking advantage of integral field spectroscopy data \citep[e.g.][]{igle,duar}. However, because we compare our data with the data from the MPA-JHU catalogue, we decided to use its aperture corrections in order to be consistent with their procedure. Because the aperture correction is a factor for each galaxy, the SFR$_{fib}$ estimates with FADO and MPA-JHU are equally affected. This means that our final goal of comparing the SFR$_{tot}$ between FADO and MPA-JHU is independent of the applied aperture correction.

\begin{figure*}[h!]
        \centering
        \includegraphics[scale=0.55]{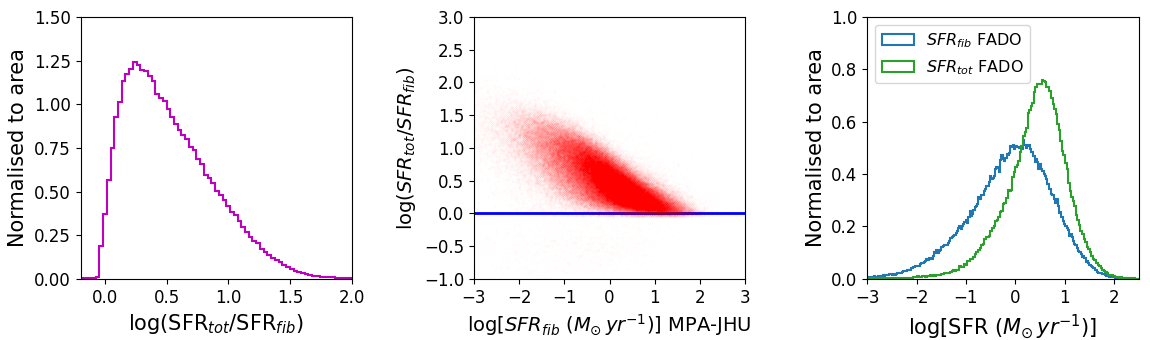}
        \caption{Conversion from fibre into total SFR for the common SF sample. Left panel: Histogram of the logarithm of the ratio of the fibre and total SFR for MPA-JHU. Middle panel: Scatter plot of the logarithm of the ratio of the fibre and total SFR as a function of the logarithm of the fibre SFR for MPA-JHU. Right panel: Histogram of the fibre SFR (blue) and total SFR (green) for FADO.}
        \label{fig:conversion_fibre_total_SFR}
    \end{figure*}

With the aim of applying the described correction to our SFR$_{fib}$ estimates, we calculated the ratio of SFR$_{tot}$ and SFR$_{fib}$ from the MPA-JHU catalogue. This ratio represents the relation between the SFR$_{tot}$ and the SFR$_{fib}$ for each galaxy, so that by multiplying this ratio by the estimated SFR$_{fib}$ , we obtained an estimate of the SFR$_{tot}$. The distribution of the ratio of the SFR$_{tot}$ and SFR$_{fib}$ from the MPA-JHU catalogue for our SF sample is presented in \autoref{fig:conversion_fibre_total_SFR}. We also show the effect of the correction on the SFR$_{fib}$ estimated with FADO. The effect of this correction in the SFR$_{fib}$ estimates for MPA-JHU is not shown since it is the same correction factor as for the FADO estimates.

The distribution of the SFR$_{fib}$ to SFR$_{tot}$ conversion factor (left panel of \autoref{fig:conversion_fibre_total_SFR}) gives an insight into the relation between the two estimates. The most common conversion factor is $\log$(SFR$_{tot}$/SFR$_{fib}$) $\approx$ 0.3, which means that the SFR$_{tot}$ is twice as high as the SFR$_{fib}$. After this peak, the distribution steadily decreases to $\log$(SFR$_{tot}$/SFR$_{fib}$) $\approx$ 2, after which there are practically no galaxies. The comparison of the SFR$_{fib}$ and SFR$_{tot}$ estimates for FADO (right panel of \autoref{fig:conversion_fibre_total_SFR}) shows that the distribution of the latter is narrower, covering one order of magnitude less at low values, but it maintains approximately the same upper limit. Moreover, the peak of the SFR$_{tot}$ distribution occurs at higher values. These facts lead to the interpretation that the galaxies with lower SFR$_{fib}$  are most strongly affected by the aperture correction. On the other hand, the galaxies with the highest SFR$_{fib}$ are almost not affected by the correction, as confirmed by the middle panel of \autoref{fig:conversion_fibre_total_SFR}.

The final SFR$_{tot}$ estimates for FADO and MPA-JHU considered in this work were obtained applying the above correction. Hereafter, only the SFR$_{tot}$ is considered, and so it is simply referred to as the SFR.

\section{Stellar mass determination}
\label{Mstar_det}

The stellar masses available in the MPA-JHU catalogue were obtained following a similar philosophy as was used in \cite{kauff}. \cite{kauff} performed a fit to the SDSS spectra following various models from \cite{bruz03}. In these fits, two spectral indices were considered: the 4000\AA\, break strength, and the Balmer absorption-line index H$\delta_{A}$. From these indices, the stellar mass-to-light ratio was estimated, which was then used to obtain the stellar mass. The stellar masses available in the MPA-JHU catalogue differs from this method as the fits are based on the broadband $u$, $g$, $r$, $i$ , and $z$ photometry. However, the differences between the two methods are very small (median difference of 0.01 dex), and they are discussed on the MPA-JHU website.

\begin{figure*}[h!]
        \centering
        \includegraphics[scale=0.55]{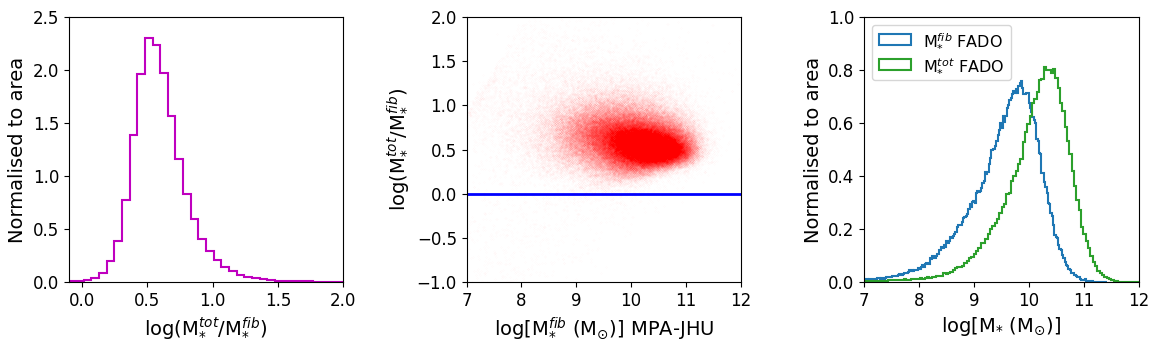}
        \caption{Conversion from fibre to total stellar mass for the common SF sample. Left panel: Histogram of the logarithm of the ratio of the fibre and total stellar mass for MPA-JHU. Middle panel: Scatter plot of the logarithm of the ratio of the fibre and total stellar mass as a function of the logarithm of the fibre stellar mass for MPA-JHU. Right panel: Histogram of the fibre stellar mass (blue) and total stellar mass (green) for FADO.}
        \label{fig:conversion_fibre_total_Mstar}
    \end{figure*}

For FADO, the stellar mass is derived through a fit to the spectrum. The stellar mass has to be corrected for aperture effects as well. We followed the same procedure as for the SFR, calculating the ratio of the total and fibre stellar mass for the MPA-JHU data and then applying this factor to the fibre stellar estimate of FADO in order to obtain the total stellar mass. The distribution of the ratio of total and fibre stellar mass from the MPA-JHU catalogue is presented in \autoref{fig:conversion_fibre_total_Mstar}. We also show the effect of the correction on the stellar mass estimated with FADO.

The distribution of the conversion factor between fibre and total stellar mass shows that the most common difference between these quantities is $\approx$0.5 dex, which corresponds to a stellar mass about three times higher than the fibre stellar mass. After the peak, the distribution decreases to a difference between total and fibre stellar mass of $\approx$1.5 dex. This correction affects the fibre stellar mass distribution by shifting it to higher values, maintaining approximately the overall shape. Hence, it can be said that this conversion factor affects most of the galaxies in approximately the same way. With the total stellar mass estimates for FADO obtained, the MPA-JHU estimates were converted from a \cite{krou} in to a \cite{chabr} IMF by dividing the estimates by 1.06 \citep{zah,spea,madau}.

\section{Results}
\label{results}

\subsection{Star formation rate}

\begin{figure*}[h!]
        \centering
        \includegraphics[scale=0.55]{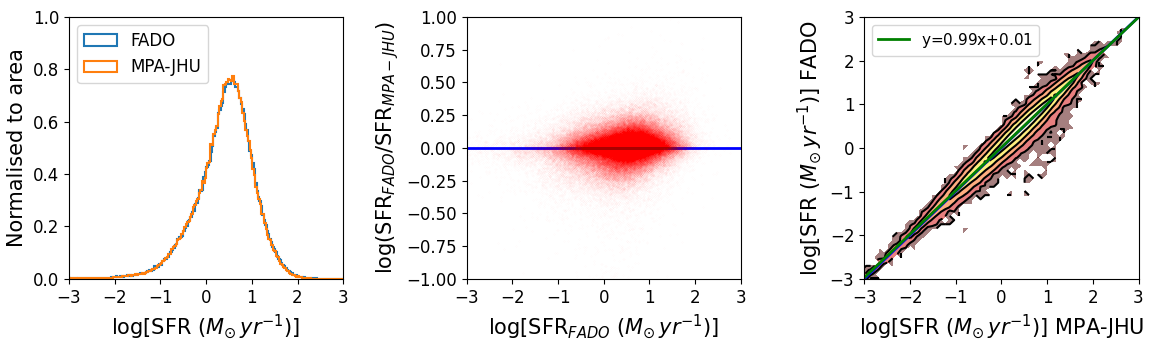}
        \caption{Comparison of the SFR between FADO and MPA-JHU. Left panel: Histogram of $\log$(SFR) for FADO (blue) and MPA-JHU (orange). Middle panel: Scatter plot of the logarithm of the ratio of the SFR from FADO and MPA-JHU as a function of $\log$(SFR) for FADO. Right panel: 1:1 contour plot showing 20, 40, 60, 80, and 100\% of the sample and linear fit to the data (green line).}
        \label{fig:total_SFR_FADO_MPA}
    \end{figure*}

The comparison between the SFR calculated from the H$\alpha$ flux from FADO and from MPA-JHU is presented in \autoref{fig:total_SFR_FADO_MPA}. The SFR distributions for FADO and MPA-JHU are coincident, which was expected because the H$\alpha$ flux distributions are also coincident. The 1:1 plots also show that FADO and MPA-JHU obtain similar SFRs. In particular, the linear fit applied to the comparison between the two estimates is practically parallel to the 1:1 reference line. This fact indicates that the relation between the two estimates remains the same in the considered range, that is, there is no change in the difference between the SFR estimate by FADO and MPA-JHU with increasing SFR. The difference between the SFR estimate of FADO and MPA-JHU differs by only 0.01 dex ($\approx$2\%).

In order to quantify the SFR estimates of FADO in more detail, the sample was divided into three subsamples: galaxies forming less than 1 M$_{\odot}$yr$^{-1}$, $\log$(SFR [M$_{\odot}$yr$^{-1}$]) $<$ 0, galaxies forming between 1 and 10 M$_{\odot}$yr$^{-1}$, 0 $\le$ log(SFR [M$_{\odot}$yr$^{-1}$]) $<$ 1, and galaxies forming more than 10 M$_{\odot}$yr$^{-1}$, $\log$(SFR [M$_{\odot}$yr$^{-1}$]) $\ge$ 1. For each subsample, we computed the number of galaxies, the median value, and the standard deviation of the $\log$(SFR). In \autoref{tab:SFR_comparison} we present the results of the analysis of these subsamples as well as the full sample. This analysis provides insight into the star formation level of the galaxies in our sample.

In the complete sample, FADO estimates a median $\log$(SFR) of 0.48 ($\approx$3 M$_{\odot}$yr$^{-1}$). The 0 $\le$ $\log$(SFR) $<$ 1 subsample contains $\approx$62\% of the complete SF sample, and the subsample with the highest SFRs, $\log$(SFR) $\ge$ 1, has the fewest galaxies ($\approx$15\%), with a median value $\log$(SFR) of 1.21 ($\approx$16 M$_{\odot}$yr$^{-1}$). This SFR characterisation shows that the galaxies in our sample do not have a very high star-formation level in general. Consequently, the nebular emission of these galaxies should also be low. Thus, our sample is likely dominated by galaxies with a low nebular contribution.

    \begin{table}[h!]
        \centering
        \caption{Analysis of the SFR, in $\log$(SFR [M$_{\odot}$yr$^{-1}$]), estimated with FADO. We present the number of galaxies, the median value, and the standard deviation of the three subsamples and the complete SF sample.}
        \label{tab:SFR_comparison}
        \begin{tabular}{ccc}
        \hline
        Sample                   & No. Galaxies & Median $\pm$ $\sigma$ \\ \hline
        $\log$(SFR) $<$ 0        & 40 230       & -0.35$\pm$0.51   \\
        0 $\le$ $\log$(SFR) $<$ 1 & 114 530      & 0.52$\pm$0.27    \\
        $\log$(SFR) $\ge$ 1      & 28 173       & 1.21$\pm$0.24    \\
        SF Sample                & 182 933      & 0.48$\pm$0.64    \\ \hline
        \end{tabular}
    \end{table}

\subsection{Stellar mass}

\begin{figure*}[h!]
        \centering
        \includegraphics[scale=0.55]{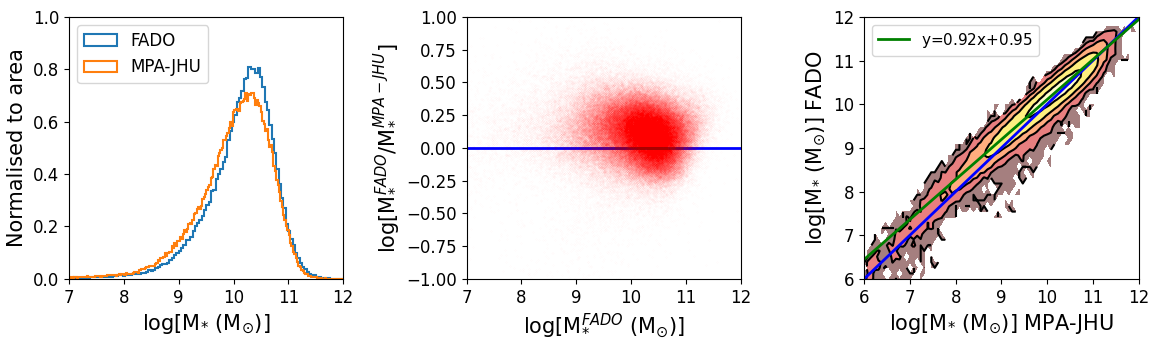}
        \caption{Comparison of the total stellar mass between FADO and MPA-JHU. Left panel: Histogram of the logarithm of the stellar mass for FADO (blue) and MPA-JHU (orange). Middle panel: Scatter plot of the logarithm of the ratio of the stellar mass from FADO and MPA-JHU as a function of the logarithm of the stellar mass. Right panel:  1:1 contour plot showing the 20, 40, 60, 80, and 100\% of the sample and linear fit to the data (green line).}
        \label{fig:Mstar_FADO_MPA}
    \end{figure*}

\autoref{fig:Mstar_FADO_MPA} compares the stellar mass for FADO and MPA-JHU. The two distributions have a similar shape, but  the peak occurs at higher stellar mass values for FADO than for MPA-JHU. This fact means that FADO retrieves slightly higher stellar masses than MPA-JHU on average, as is also shown in the middle and right panel of \autoref{fig:Mstar_FADO_MPA}. However, as the stellar mass increases, the difference between FADO and MPA-JHU decreases. This can also be seen for high masses, where the distributions are consistent.

The complete sample was divided into two subsamples based on the stellar mass: galaxies with a stellar mass lower and higher than 10$^{10}$ M$_{\odot}$.  \autoref{tab:Mass_comparison} presents the results of the analysis.

The uncertainty presented for the FADO estimates (0.2 dex) is a conservative upper limit error for the stellar mass obtained in previous studies of FADO \citep{gom,cardo,pappa}. The uncertainties given directly by FADO for the stellar mass estimates are the formal errors from the minimisation algorithm, related to the error of the $\chi^{2}$ distribution, and not the physical error affecting the stellar mass estimate. Since these formal errors were unrealistically low, we decided to fix the value of 0.2 dex, based on the aforementioned previous assessments. For MPA-JHU, the uncertainties presented are the 16th and 84th percentile of their stellar mass estimate. The uncertainty of the median difference between FADO and MPA-JHU is the result of adding in quadrature the FADO and MPA-JHU uncertainties.

For the complete sample, the median difference between FADO and MPA-JHU is 0.11 dex. This means that in median, FADO estimates stellar masses $\text{that are about }$29\% higher than MPA-JHU. For both subsamples, the median values of FADO are higher than those of MPA-JHU. The median difference between FADO and MPA-JHU for the subsample of stellar masses lower and higher than 10$^{10}$ M$_{\odot}$ is 0.14 dex and 0.10 dex, respectively. These results are consistent with the trend observed in the linear fit to the 1:1 contour plot (right panel of \autoref{fig:Mstar_FADO_MPA}) of a decrease in the difference between FADO and MPA-JHU with increasing stellar mass.

To test the nature of the origin of the observed differences, we applied the statistical Kolmogorov-Smirnov test and obtained a p-value of $\sim$0. This value is lower than any of the commonly used confidence levels. The hypothesis that the distributions are similar in statistical terms can therefore be rejected. However, the measured difference between the stellar masses derived by FADO and MPA-JHU are in any case within 1$\sigma$ of the uncertainties for each method. This implies that even if the distributions are statistically different, the uncertainties associated with the measurements dominate and the compatibility of the two estimates cannot be ruled out.

This is also true for the observed trend of a decrease in the difference between FADO and MPA-JHU with increasing stellar mass, and an interpretation of this trend is not statistically significant. It is worth mentioning that although the stellar masses are derived through different approaches in FADO and MPA-JHU, as stated in Section \ref{Mstar_det}, the estimates still agree.

    \begin{table*}[h!]
        \centering
        \caption{Analysis of the stellar mass, in $\log$(M$_{*}$ [M$_{\odot}$]), for FADO and MPA-JHU. We present the number of galaxies, the median value for FADO and MPA-JHU, and the median difference between FADO and MPA-JHU for two subsamples and the complete sample. The uncertainty associated with FADO is a conservative upper limit, and for MPA-JHU, it is the 16th and 84th percentile of the estimate. The uncertainty associated with the median difference is the result of adding the FADO and MPA-JHU uncertainties in quadrature.}
        \label{tab:Mass_comparison}
        \resizebox{15 cm}{!}{%
        \begin{tabular}{ccccccc}
        \hline
         & \multicolumn{2}{c}{$\log$(M$_{*}$) $<$ 10} & \multicolumn{2}{c}{$\log$(M$_{*}$) $\ge$ 10} & \multicolumn{2}{c}{Complete Sample} \\
         & Nº Gal & Median & No. Gal & Median & No. Gal & Median \\ \hline
         FADO & \multirow{3}{*}{63 412} & 9.6$\pm$0.2 & \multirow{3}{*}{119 521} & 10.4$\pm$0.2 & \multirow{3}{*}{182 933} & 10.2$\pm$0.2 \\
         MPA-JHU & & 9.45$^{+0.09}_{-0.07}$ & & 10.36$^{+0.10}_{-0.09}$ & & 10.12$^{+0.10}_{-0.08}$ \\
         FADO$-$MPA-JHU & & 0.14$^{+0.22}_{-0.21}$ & & 0.10$^{+0.22}_{-0.22}$ & & 0.11$^{+0.22}_{-0.22}$ \\ \hline
        \end{tabular}
        }
    \end{table*}

\subsection{Main sequence}

We showed and quantified that FADO obtains similar SFRs and stellar masses as MPA-JHU. Since the SFMS is the combination of these two galactic properties, small differences are expected between the SFMSs obtained using the results from FADO and MPA-JHU.

\begin{figure*}[h!]
        \centering
        \includegraphics[scale=0.6]{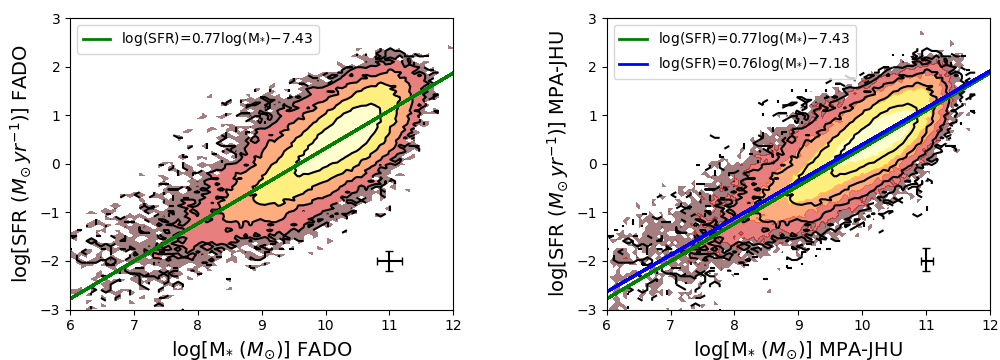}
        \caption{Comparison of the star-forming main sequence with FADO and MPA-JHU data. Left panel: Contour plot for FADO and linear fit to the data (green line). Right panel: Contour plot for MPA-JHU (colours) and linear fit to the data (blue line) with the contours (black lines) and linear fit (green line) for FADO overplotted. The median error of the data is presented in the bottom right corner, and the linear fit equation is presented in the upper left corner. The contour plots show 20, 40, 60, 80, and 100\% of the sample.}
        \label{fig:Main_sequence}
    \end{figure*}

The SFMS obtained with FADO and MPA-JHU data are plotted in \autoref{fig:Main_sequence}. A linear fit is applied to the SFMS, and the resulting equation presented in the plot. The contours and linear fit of the SFMS for FADO are overplotted on the SFMS for MPA-JHU in order to easily compare both. We also present a cross with the median error affecting the SFR and stellar mass estimates.

The obtained equations for the linear fits, as well as the formal errors from the fitting, are

    \begin{equation}
     \begin{aligned}
        \log(\textrm{SFR}) &= (0.774\pm0.001)\log(M_{*})-(7.43\pm0.02) \qquad \mbox{FADO} \\
        \log(\textrm{SFR}) &= (0.757\pm0.001)\log(M_{*})-(7.18\pm0.01) \ \mbox{MPA-JHU}
     \end{aligned}
    .\end{equation}

As expected, the SFMSs are consistent with each other, and the linear fits applied to the data present the same relation between SFR and stellar mass. Taking into account the error cross presented in the plot, we conclude that the linear fits are not statistically different from each other. Moreover, the contour plot shows that the distribution of the galaxies within the SFMS is similar for FADO and MPA-JHU.

\section{Discussion}
\label{discussion}

As discussed in Section \ref{method}, the main novelty of FADO from the astrophysical point of view is the inclusion of the nebular component in the fitting process of the SED. The overall continuum of the FADO model SED is therefore the result of the contribution from the stellar and nebular continuum. On the other hand, previous spectral synthesis methods, such as the one used to estimate the physical parameters in the MPA-JHU catalogue, only consider the stellar contribution. Therefore, the overall continuum is assumed to be purely stellar. 

The impact of considering the nebular contribution in the fitting process of the SED was recently studied in \cite{cardo_prep} and \cite{breda}. In both these works, FADO, with its self-consistent treatment of the stellar and nebular continuum, was compared to the purely stellar synthesis code STARLIGHT. Because these works addressed a similar topic as ours, it is worth analysing our results in the context of their findings.

\cite{cardo_prep} reevaluated the evolutionary trend of light- and mass-weighted quantities related to the stellar age, metallicity, and mass of  SF galaxies in light of the new approach followed by FADO. This comparison was carried out considering different subsamples, defined in terms of the star formation level of the galaxies using the EW as indicator. In this way, the stellar properties of the galaxies in the different subsamples were related to the level of the nebular continuum contribution. Of the different results presented in \cite{cardo_prep}, the most relevant result for this work is that the estimated stellar mass is minimally affected by taking into consideration the nebular continuum because FADO obtains stellar masses just slightly higher than STARLIGHT \citep[see Table 1 of][]{cardo_prep}.

 \cite{breda} studied a sample of 414 extreme emission line galaxies (EELGs) and 697 normal SF galaxies from the SDSS, providing insight into how stellar properties, such as stellar mass and mean stellar age, are affected by the additional modelling of the nebular continuum. The specific SFR (sSFR) and the sum of the flux of the most prominent emission lines were used as indicators of the degree of star formation activity, thus enabling the interpretation of the results considering the level of the nebular contribution. In EELGs, STARLIGHT estimates stellar masses higher than FADO as a consequence of the increased level of nebular emission. Normal SF galaxies are only a slightly overestimated by STARLIGHT, however.

As stated in Section \ref{method}, we considered a specific subsample of the application of FADO to SDSS-DR7 from \cite{cardo_prep}. We obtained that the H$\alpha$ flux estimates of FADO and MPA-JHU are consistent with each other. The consistency between the fluxes estimated by FADO and those in the MPA-JHU catalogue is a test of the reliability of the flux estimates from FADO. Furthermore, our results show that FADO estimates H$\alpha$ EWs that are $\approx26\%$ higher than the values in the MPA-JHU catalogue. This is expected because FADO uses the monochromatic continuum level of the SED at the central wavelength of emission lines to determine the EW, whereas the approach followed in MPA-JHU considers the continuum as the average level of the continuum adjacent to the emission or absorption line (see Section \ref{lines} for further details).

Our comparison of the stellar mass between FADO and MPA-JHU showed that the median values are similar considering the uncertainty range, but the distributions are slightly different. The estimate of comparable stellar masses by FADO relative to a purely stellar approach to SED fitting was also observed in \cite{cardo_prep}. However, \cite{breda} observed that not considering the nebular contribution leads to an overestimation of the stellar mass.

Since FADO considers the contribution of the stellar and nebular component to the total continuum, but a purely stellar approach only the stellar component, then the stellar continuum for FADO is at a lower level. Consequently, the stellar mass estimates of FADO are expected to be lower than those of a pure stellar code. However, this difference will only be significant for galaxies for which the contribution of the nebular continuum to the SED becomes significant.

In this way, the apparent disagreement of the results from \cite{breda} with the results reported here and in \cite{cardo_prep} is probably related to the level of the nebular continuum contribution of the galaxies that is included in the sample used in each of these works. In our sample, the median value of the H$\alpha$ EW is 34 \AA\, and only 6\% of the galaxies in the sample have an EW(H$\alpha$)>100 \AA\, (in \autoref{fig:observed_EW} is presented the EW(H$\alpha$) distribution). In \cite{cardo_prep}, only $\approx1.3\%$ of the galaxies have an EW(H$\alpha$)>75 \AA\,\citep[see Fig.2 of][for the EW(H$\alpha$) distribution]{cardo_prep}. This means that these samples are strongly dominated by galaxies with a negligible nebular continuum, which is not expected to appreciably influence purely stellar spectral fits. Thus, an agreement between the physical properties retrieved by FADO and by a pure stellar code is observed, as is the case in what regards the MPA-JHU catalogue. On the other hand, in \cite{breda} the sample of EELGs has an average H$\alpha$ EW of $\approx$400 \AA\, \citep[see Fig.1 of][for the EW(H$\alpha$) distribution]{breda}. Therefore, the nebular contribution to the galaxies is so high that it is not negligible, and in order to retrieve accurate physical properties, such as the stellar mass, it is fundamental to model the nebular continuum.

Our comparison of FADO and MPA-JHU is worth extending to EELGs. We selected a subsample of EELGs from our SF sample by adopting the same cuts in EW as  were followed in \cite{breda}. Thus, we selected the galaxies with EW(H$\beta$) $>$ 30 \AA\, and EW([OIII]) $>$ 100 \AA\, for both FADO and MPA-JHU. With these criteria, we obtained a subsample of 2759 EELGs, which corresponds to $\approx$ 1.5\% of our SF sample. The EELGs sample has a median H$\alpha$ EW of 316 \AA\,, while for the SF sample the median H$\alpha$ EW is 34 \AA.

\begin{figure}[h!]
        \centering
        \includegraphics[scale=0.5]{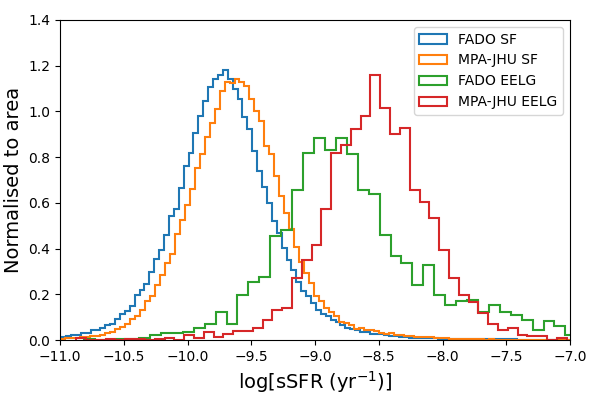}
        \caption{Histograms of the logarithmic sSFR of the SF galaxies and EELG samples for FADO (in blue and green, respectively) and MPA-JHU (in orange and red, respectively).}
        \label{fig:sSFR_EELGs_nSF}
    \end{figure}

In order to compare the EELGs sample with the complete SF sample, the sSFR of these two samples for FADO and MPA-JHU are presented in \autoref{fig:sSFR_EELGs_nSF}. There is a clear distinction between the EELGs and normal SF galaxies. The peak in the distributions of the EELGs is substantially shifted towards higher values, which means that the EELGs generally have higher sSFR than normal SF galaxies.

The sSFR for MPA-JHU is higher than for FADO in both samples. This is a consequence of the difference in the stellar mass estimates between FADO and MPA-JHU because the SFR estimates are similar between the two. However, the difference between FADO and MPA-JHU is higher for the EELGs than for the normal SF galaxies. This trend is opposite to the results in \cite{breda}, where FADO obtains even higher sSFRs for the EELGs than for the normal SF galaxies compared to the pure stellar approach. However, the uncertainties are again larger than the differences, even though the distributions are clearly different, and the compatibility of both estimates cannot be ruled out.

This difference in trend could be related to the differences in the spectral basis used in these works, both in terms of the age, in \cite{breda} the age upper limit of the basis depends on the redshift of the galaxy, and of the metallicity, in \cite{breda} the basis spans a smaller range, particularly at sub-solar metallicities. These differences in the basis lead to different population vectors, thus changing the derived physical properties, including the stellar mass. Finally, even if the same spectral basis is considered, different approaches could lead to different results due to the effect of the fitting algorithm on the derived population vector \cite{cardo_prep}. Thus, the fact that in this work FADO is compared with MPA-JHU, but in \cite{breda} it is compared with STARLIGHT, and that the three approaches use different fitting algorithms could lead to relevant differences.

These points underline the difficulty of comparing different approaches to SED fitting because the chosen spectral basis and the employed fitting algorithms play a non-negligible role. Another relevant point to take into consideration in future works is to refine the error estimates for these physical parameters so that the effect of the nebular contribution can be better quantified.

In summary, our results show that the stellar mass is consistent between FADO and MPA-JHU because the vast majority of SF galaxies in our sample have a low nebular contribution. Thus, the modelling or lack thereof of the nebular contribution does not affect the retrieved physical properties of these galaxies. However, there is a subsample of SDSS galaxies with a significant high nebular contribution, such as the EELGs studied in \cite{breda} or luminous compact galaxies \citep[e.g.][]{izo11,amorin,papad12}, for which the modelling of the nebular continuum is fundamental to retrieve more reliable physical parameters.

In addition to particular types of SDSS galaxies, FADO will probably be proven valuable at higher redshifts, where the galaxies have higher SFRs and nebular contribution \citep{scha,atek}. Consequently, applying FADO to galaxies at redshifts higher than the redshift coverage of SDSS will be crucial to understand the impact of the nebular contribution at these redshifts, where we expect to probe a population that is generally less evolved and more dust free. SDSS galaxies with a high nebular contribution are considered to be prototypes of higher-redshifts galaxies. Hence, results such as those presented in \cite{breda} indicate that a self-consistent treatment of the stellar and nebular continuum is required at higher redshifts in order to obtain accurate estimates of the physical properties of galaxies. The application of FADO to galaxies at higher redshifts than SDSS should also take into account that the high nebular contribution might not be the only difference relative to local galaxies because the stellar populations that compose these galaxy might also be different \citep{oli,ume}.

\section{Summary and conclusions}
\label{conclusion}

FADO is an innovative spectral synthesis code that considers both the stellar and nebular components in a self-consistent way while fitting the SED of a galaxy. We considered the application of FADO to the SDSS-DR7 carried out in \cite{cardo_prep}. Our goal was to study the impact of the nebular contribution on the estimation of the physical properties of SF main-sequence galaxies. We selected a sample of 182 933 local SF galaxies and compared the results to the well-established MPA-JHU catalogue, whose high quality and reliable results were used in previous benchmark publications.

The SFR was estimated from the flux of the H$\alpha$ emission line. Firstly, the flux and EW of H$\alpha$ were studied. FADO retrieves in median H$\alpha$ fluxes similar to MPA-JHU and EWs 0.1 dex ($\approx$26\%) higher. This difference is a consequence of a different definition of the level of the continuum between FADO and MPA-JHU. FADO uses the monochromatic continuum level of the SED at the central wavelength of emission lines to determine the EW, whereas the usual approach, which is followed in MPA-JHU, is to consider the continuum as the average level of the continuum adjacent to the emission/absorption line.

In order to correct the fiber-determined SFRs and stellar masses for aperture effects, the same aperture corrections as in MPA-JHU \citep{brinch,salim} were adopted. These corrections have a typical value of 0.3 and 0.5 dex for the SFR and stellar mass, respectively, and can reach values as high as 1 dex for both quantities. Considering the magnitude of these corrections, they are crucial to the slope and scatter of the SFMS.

Proceeding to determine the SFR, we observed an agreement between FADO and MPA-JHU, as for the H$\alpha$ flux, with a median difference between the two data sets of 0.01 dex ($\approx$3\%). In order to analyse the estimates in more detail, the sample was divided into three bins, $\log$(SFR) $<$ 0, 0 $\le$ $\log$(SFR) $<$ 1, and $\log$(SFR) $\ge$ 1. This analysis of the SFR shows that our sample is likely dominated by galaxies with a low nebular contribution (see \autoref{fig:total_SFR_FADO_MPA} and \autoref{tab:SFR_comparison} for details).

For most galaxies, the stellar masses retrieved by FADO are higher than in MPA-JHU, with a median difference of 0.11 dex ($\approx$29\%). The median difference between FADO and MPA-JHU was evaluated for two different subsample: galaxies with a stellar mass lower and higher than 10$^{10}$ M$_{\odot}$. The results show that the differences decrease as the stellar mass increases. However, the observed differences are contained within the uncertainty range, and consequently, the stellar mass estimates are consistent with each other (see \autoref{fig:Mstar_FADO_MPA} and \autoref{tab:Mass_comparison} for details).

The consistent estimates of the SFR and stellar mass between FADO and MPA-JHU lead to a similar SFMS (see \autoref{fig:Main_sequence}). From these results, we conclude that for local and normal SF galaxies, the nebular contribution does not affect the determination of the SFMS significantly because the nebular continuum is weak on average. Nonetheless, for specific SDSS galaxies, such as starburst galaxies, the nebular contribution is sufficiently high to influence the determination of the physical properties \citep{izo11,amorin,papad12,breda}.

This work shows that taking into account the contribution of the nebular component to the galaxy SED does not affect SFR tracers based on emission line fluxes. Moreover, for local and normal SF galaxies, the inclusion of the nebular component does not lead to significant differences in the stellar mass estimates in the specific context of aperture-corrected estimates. In the end, the linear relation between SFR ans stellar mass known as SFMS is not significantly affected by modelling the nebular continuum for local and normal SF galaxies.

From studies at higher redshifts, it seems clear that the nebular emission is significantly higher than in local SF galaxies. Thus, it is fundamental to study its influence on the physical properties of galaxies at higher redshifts with FADO and other SED fitting codes. This line of research is particularly relevant in order to understand the most reliable tool for modelling and analysing the data from future high-redshift surveys, such as MOONS \citep{cira16,cira20,mai}, PFS \citep{taka}, WEAVE \citep{dalt}, and 4MOST \citep{dejong}.

\begin{acknowledgements}

This work was supported by Fundação para a Ciência e a Tecnologia (FCT) through the research grants PTDC/FIS-AST/29245/2017, UID/FIS/04434/2019, UIDB/04434/2020 and UIDP/04434/2020. C. P. acknowledges support from DL 57/2016 (P2460) from the ‘Departamento de Física, Faculdade de Ciências da Universidade de Lisboa’. I. M. acknowledges support from DL 57/2016 (P2461) from the ‘Departamento de Física, Faculdade de Ciências da Universidade de Lisboa’. APA acknowledges support from the Fundação para a Ciência e a Tecnologia (FCT) through the work Contract No. 2020.03946.CEECIND, and through the FCT project EXPL/FIS-AST/1085/2021. R.C. acknowledges support from the Fundação para a Ciência e a Tecnologia (FCT) through the Fellowship PD/BD/150455/2019 (PhD:SPACE Doctoral Network PD/00040/2012). Funding for the SDSS and SDSS-II has been provided by the Alfred P. Sloan Foundation, the Participating Institutions, the National Science Foundation, the U.S. Department of Energy, the National Aeronautics and Space Administration, the Japanese Monbukagakusho, the Max Planck Society, and the Higher Education Funding Council for England. The SDSS Web Site is \url{http://www.sdss.org/}. The SDSS is managed by the Astrophysical Research Consortium for the Participating Institutions. The Participating Institutions are the American Museum of Natural History, Astrophysical Institute Potsdam, University of Basel, University of Cambridge, Case Western Reserve University, University of Chicago, Drexel University, Fermilab, the Institute for Advanced Study, the Japan Participation Group, Johns Hopkins University, the Joint Institute for Nuclear Astrophysics, the Kavli Institute for Particle Astrophysics and Cosmology, the Korean Scientist Group, the Chinese Academy of Sciences (LAMOST), Los Alamos National Laboratory, the Max-Planck-Institute for Astronomy (MPIA), the Max-Planck-Institute for Astrophysics (MPA), New Mexico State University, Ohio State University, University of Pittsburgh, University of Portsmouth, Princeton University, the United States Naval Observatory, and the University of Washington.

\end{acknowledgements}

\begin{appendix}
\section{Extinction correction}
\label{appendix}

The effect of extinction $A(\lambda)$ can be quantified as the difference between the observed magnitude $m(\lambda)$ and the magnitude in the absence of dust and gas $m_{0}(\lambda)$,

    \begin{equation}
        m(\lambda)-m_{0}(\lambda) = A(\lambda) 
        \label{eq:extinction}
    .\end{equation}

\noindent Following the definition of magnitude, the difference between $m(\lambda)$ and $m_{0}(\lambda)$ is related to the observed flux $F(\lambda)$ and the flux in the absence of dust $F_{0}(\lambda)$ by

    \begin{equation}
        m(\lambda)-m_{0}(\lambda) = -2.5\log \left( \dfrac{F(\lambda)}{F_{0}(\lambda)} \right) 
        \label{eq:magnitude}
    .\end{equation}

\noindent By joining \autoref{eq:extinction} and \autoref{eq:magnitude}, a relation between $F_{0}(\lambda)$ and $F(\lambda)$ as a function of $A(\lambda)$ can be derived,

    \begin{equation}
        A(\lambda) = -2.5\log \left( \dfrac{F(\lambda)}{F_{0}(\lambda)} \right) \Leftrightarrow F_{0}(\lambda) = F(\lambda)10^{0.4 A(\lambda)}
        \label{eq:flux}
    .\end{equation}

\noindent  \autoref{eq:flux} is only valid for an uniform layer of dust between the source and the observer, that is, assuming that the extinction does not depend on the geometry.

Dust does not affect the different wavelengths of the passing radiation equally. The flux is more diminished for smaller wavelengths than for higher wavelengths. In this way, to characterise the dust extinction, we compared the flux from two different wavelengths since they are differently affected by the presence of dust. It is reasonable to choose lines characteristic of regions where the nebular lines are produced, thus we used the hydrogen emission lines H$\alpha$ and $H\beta$. The ratio of the flux of H$\alpha$ and $H\beta$ is known as the Balmer decrement, and it was used to correct the observed fluxes. The difference between the extinction at $H\beta$ and H$\alpha$ is

    \begin{equation}
            A(H\beta)-A(H\alpha) = 2.5 \log \left( \dfrac{(H\alpha/H\beta)_{obs}}{(H\alpha/H\beta)_{int}} \right)
        \label{eq:Hb_Ha_dif}
    ,\end{equation}

\noindent where (H$\alpha$/H$\beta$)$_{obs}$ and (H$\alpha$/H$\beta$)$_{int}$ are the observed and intrinsic Balmer decrement, respectively. Assuming an electron density n = 100 cm$^{-3}$ and electron temperature T = 10$^{4}$ K for case-B recombination, then (H$\alpha$/H$\beta$)$_{int}$ = 2.86 \citep{oster}.

With \autoref{eq:Hb_Ha_dif}, the difference between the extinction at $H\beta$ and H$\alpha$ can be calculated, which is called colour excess. The broadband colour excess $E(B-V)$ and an extinction law $k(\lambda)$ can be used to calculate the extinction at any wavelength, which is the final objective, via

    \begin{equation}
        A(\lambda)=k(\lambda) E(B-V) 
        \label{eq:broadband_colour_excess}
    .\end{equation}

\noindent The previous equation can be used to write the difference between the extinction at H$\beta$ and H$\alpha$,

    \begin{equation}
        A(H\beta)-A(H\alpha) = [k(H\beta)-k(H\alpha)] E(B-V)
        \label{eq:colour_excess_extinction}
    ,\end{equation}

\noindent where $k($H$\beta)$ and $k($H$\alpha)$ are the values of the extinction law at H$\beta$ and H$\alpha$, respectively. Both \autoref{eq:Hb_Ha_dif} and \autoref{eq:colour_excess_extinction} provide a way in which the difference between the extinction at H$\beta$ and H$\alpha$ can be calculated. These two equations together read 

    \begin{equation}
        E(B-V) = \dfrac{2.5}{k(H\beta)-k(H\alpha)} \log \left( \dfrac{(H\alpha/H\beta)_{obs}}{(H\alpha/H\beta)_{int}} \right) 
        \label{eq:colour_excess_Balmer_decrement}
    .\end{equation}

\noindent In this way, the broadband colour excess can be derived from the observed Balmer decrement.

Lastly, it is necessary to define an extinction law. We used the extinction law from \cite{char_fall}, which  is given by
    \begin{equation}
        k_{Char}(\lambda) = \left( \dfrac{\lambda}{5500} \right)^{-0.7}
    ,\end{equation}

\noindent where $\lambda$ is in Ångströms. Using this extinction law, we obtain  $k_{Char}(H\alpha) \simeq 0.88$ and $k_{Char}(H\beta) \simeq 1.09$. 

Summarising, to correct the observed flux for the nebular extinction, the procedure we followed was

    \begin{enumerate}
        \item Calculate the ratio of the extinction at H$\beta$ and H$\alpha$ with \autoref{eq:Hb_Ha_dif}.
        
        \item Calculate the broadband colour excess $E(B-V)$ with \autoref{eq:colour_excess_Balmer_decrement}.
    
        \item Calculate the extinction $A(\lambda)$ at the desired wavelength with \autoref{eq:broadband_colour_excess}.
    
        \item Calculate the flux in the absence of dust $F_{0}(\lambda)$ with \autoref{eq:flux}.
    \end{enumerate}

\end{appendix}
\end{document}